\documentclass[12pt]{article}

\usepackage{amsfonts,amsmath}
\usepackage{latexsym}

\usepackage{epsfig}
\usepackage{psfrag}

\usepackage{epsfig}
\def\void{}
\def\labelmark{}

\newenvironment{formula}[1]{\def\labelname{#1}
\ifx\void\labelname\def\junk{\begin{displaymath}}
\else\def\junk{\begin{equation}\label{\labelname}}\fi\junk}%
{\ifx\void\labelname\def\junk{\end{displaymath}}
\else\def\junk{\end{equation}}\fi\junk\labelmark\def\labelname{}}

{\ifx\void\labelname\def\junk{\end{array}\end{displaymath}}
\else\def\junk{\end{array}\right.\end{equation}}
\fi\junk\labelmark\def\labelname{}\def\junk{}
}

\newcommand{\beq}{\begin{formula}}
\newcommand{\eeq}{\end{formula}}
\newcommand{\beqv}{\begin{formula}{}}

\newcommand{\rf}[1]{(\ref{#1})}
\newcommand{\oh}{\frac{1}{2}}

\newcommand{\bea}{\begin{eqnarray}}
\newcommand{\eea}{\end{eqnarray}}
\newcommand{\beas}{\begin{eqnarray*}}
\newcommand{\eeas}{\end{eqnarray*}}
\newcommand{\beqs}{\begin{displaymath}}
\newcommand{\eeqs}{\end{displaymath}}

\newcommand{\ben}{\begin{equation}}
\newcommand{\een}{\end{equation}}

\newcommand{\bdm}{\begin{displaymath}}
\newcommand{\edm}{\end{displaymath}}

\newcommand{\bbN}{\mathbb N}

\begin{document}
 \topmargin 0pt
 \oddsidemargin 5mm
 \headheight 0pt
 \topskip 0mm

 \addtolength{\baselineskip}{0.5\baselineskip}

\hfill

\vspace{1cm}

\begin{center}

{\Large \bf The spectral dimension of generic trees}

\medskip

\vspace{1.5 truecm}

{\bf Bergfinnur Durhuus}\footnote{durhuus@math.ku.dk}

\vspace{0.4 truecm}

Matematisk Institut, Universitetsparken 5

2100 Copenhagen \O, Denmark

 \vspace{.7 truecm}

{\bf Thordur Jonsson}\footnote{
thjons@raunvis.hi.is}

\vspace{0.4 truecm}

University of
Iceland, Dunhaga 3

107 Reykjavik, Iceland

 \vspace{.7 truecm}

{\bf John F. Wheater}\footnote{j.wheater1@physics.ox.ac.uk}

\vspace{.4 truecm}  

Rudolf Peierls Centre for
Theoretical Physics, University of Oxford 

1 Keble Road, OX13NP, UK

 \vspace{1.3 truecm}

 \end{center}

 \noindent
 {\bf Abstract.} We define generic ensembles of infinite trees. 
These are limits as $N\to\infty$ of ensembles of
finite trees of fixed size $N$, defined in terms of a set of branching
weights. Among these ensembles
are those supported on trees with vertices of a uniformly bounded order. 
The associated probability measures
are supported on trees with
a single spine and Hausdorff dimension $d_h
=2$. Our main result is that their spectral dimension is $d_s=4/3$, and
that the critical exponent of the mass, defined as the exponential
decay rate of the two-point function along the spine, is $1/3$.

 \newpage
 \pagestyle{plain}

\section{Introduction}
Diffusion on random geometric structures has received considerable
attention in recent years. The motivation comes from a wide range of
different areas of physics such as: percolation theory where the
 percolation clusters provide fluctuating geometries;
the physics of random media, where the effect of
impurities is often modelled by a random geometry, see e.g.\ \cite{bookben};
and
quantum gravity, where space-time itself is treated as a fluctuating
manifold, see e.g.\ \cite{book}.
 In particular, the long time characteristics of diffusion have been
studied for the purpose of providing quantitative information on the
mean large scale behavior of the geometric objects in question.  The
{\it spectral dimension} is one of the simplest quantities which provides
such information. 

In this
article the geometric structures under consideration are tree graphs
with a distinguished vertex $r$, called the root.  
The spectral dimension
$d_s$ is given by
\beq{specdim0}
p(t) \sim t^{-d_s/2}\quad \text{for $t\to\infty$}\,,
\eeq
where $p(t)$ denotes the return probability for a simple random walk
starting at $r$ as a function of (discrete) time $t$, averaged with
respect to the given probability distribution of graphs. 
In Section 4
below we calculate $d_s$ in terms of the singularities of the 
generating function for the
sequence $p(t),\,t\in \mathbb N$. 
 
 There is another natural                     
  notion of  dimension for random geometries, the 
  {\it Hausdorff dimension} $d_h$,
defined as               
\beq{haus}               
 V(R)\sim R^{d_h},               
 \eeq                
 where $V(R)$ is the ensemble average of the volume of a ball of radius $R$. 
In general it is easier to evaluate the Hausdorff dimension and we will see
that it is $2$ for all the ensembles studied in this paper.  For fixed graphs
the spectral and Hausdorff dimensions are related by
\beq{relation}
d_h\geq d_s\geq {2d_h\over 1+d_h},
\eeq
provided both exist \cite{coulhon}.  
This relation is also satisfied in some examples of
random geometries \cite{DJW}.
 
The exact value of $d_s$ is only known for a rather limited class of
models. For bond percolation on a hypercubic lattice the
value of $d_s$ for the incipient infinite cluster at criticality is
unknown, but it is conjectured to be $4/3$ in 
sufficiently high dimensions \cite{alexander}. 
For planar random surfaces
related to two-dimensional quantum gravity the spectral dimension 
is likewise unknown, but
conjectured on the basis of numerical simulations and scaling relations 
to be $2$ \cite{earlynumsim,numerics,scaling}.  For recent simulations 
investigating 
the spectral dimension in higher dimensional gravity, see
\cite{specdim1,specdim2}.

In a preceding  
article \cite{DJW} we developed techniques for analysing a
particular class 
of random geometries, called {\it random combs}, which are special
tree graphs composed of an infinite linear chain, called the {\it
spine}, to which a number of linear chains, called {\it teeth}, are
attached according to some probability distribution. In particular, we
determined the spectral dimension as well as other critical exponents
for various random combs. The techniques of \cite{DJW}, however, are 
not strong enough to
deal with general models of random trees, not to mention other
models of random graphs. The main purpose of this article is to reinforce
these methods thus enabling us to
determine the spectral dimension of a large class of random  
tree models.   This is the class
of {\it generic infinite tree ensembles} which we define in the 
next section. 
Among these ensembles
are those supported on trees with vertices of a uniformly bounded order.
Our main result
is the following.

\medskip
\noindent
{\bf Theorem 1.}
{\it The spectral dimension of generic infinite tree ensembles is

\beq{4/3}
d_s = 4/3\,.
\eeq
}

\medskip 
\noindent
Included in this class are some models that 
have been considered 
previously,
see \cite{tjjw, donetti1, jwc}. In a recent article
\cite{BarKum} it is proven for  critical
percolation on a Cayley tree that the scaling
(\ref{specdim0}) holds almost surely for individual infinite 
clusters with
$d_s=4/3$, up to logarithmic factors.
Adapting
the techniques of  \cite{BarKum}, 
similar results should be obtainable for
the models considered here. 
The results of the present paper, dealing with averaged quantities, provide
a complementary perspective.  We believe
that our method of proof is conceptually
very simple, in addition to
being applicable to a large class of random tree models.

This article is organized as follows. In Section 2 we define
the models of trees that will be considered and
describe in some detail their probability distribution, which is
supported on trees with one infinite spine. 
In Section 3 we explain the connection between our ensembles of trees and
trees that are generated by a Galton-Watson process.  This connection 
allows us to use
some well known results about such processes to analyse the trees.
Section 4 contains a proof of Theorem 1.  In Section 5 we
introduce the critical exponent of the mass and prove that it equals
$1/3$ for generic infinite trees.  This means that diffusion along the spine
of the generic infinite trees is anomalous as is discussed in detail for 
random combs in \cite{DJW}.
Finally, Section 6 contains a few concluding remarks on possible extensions
and open
problems.  Some technical results are relegated to two
appendices.

\bigskip

\section{Generic ensembles of infinite trees}

An ensemble of random graphs is a set of graphs equipped with a probability
measure.
In this section we define the ensembles of trees to
be investigated in this paper.  We start by
defining a probability measure
on finite trees and show that it yields a limiting measure on
infinite trees. 

Let $\Gamma$ be the set of all
planar rooted trees, finite or infinite, such that the root, $r$, is
of order (or valency) 1 and all vertices have finite order. 
If $\tau\in\Gamma$ is finite we let $|\tau|$
denote its size, i.e.\ the number of links in $\tau$, and the subset
of $\Gamma$ consisting of trees of fixed size $N$ will be called
$\Gamma_N$.  The subset consisting of the infinite trees will be denoted 
$\Gamma_\infty$.  Given a tree $\tau\in\Gamma$, the  
ball $B_R(\tau)$ of radius $R$ around the root is the subgraph of
$\tau$ spanned by the vertices whose graph distance from $r$ is less
than or equal to $R$. 
Note that $B_R(\tau )$ is again a rooted tree.
It is useful to define the distance
$d_{\Gamma}(\tau,\tau')$ between two 
trees $\tau, \tau'$ as $(R+1)^{-1}$, where $R$ is the
radius of the largest ball around $r$ common to $\tau$ and
$\tau'$. We shall view $\Gamma$ as a metric space with metric
$d_{\Gamma}$, see \cite{bergfinnur} for some of its properties.
For $\tau\in\Gamma$ we let $\tau\setminus r$ denote the set of all vertices 
in $\tau$ except the root.

Given a set of non-negative {\it branching weights} $w_n,\,n\in\mathbb
N$, we define the {\it finite volume partition
functions}, $Z_N$, by
\beq{x1}
Z_N = \sum_{\tau\in\Gamma_N}\prod_{i\in\tau\setminus r}w_{\sigma_i}\,,
\eeq
where $N$ is a positive integer and $\sigma_i$ denotes the
order of vertex $i$. We assume $w_1>0$, since $Z_N$ vanishes
otherwise, and we also assume $w_n>0$ for some $n\geq 3$ since
otherwise only the linear chain of length $N$ would contribute to $Z_N$.
Under these assumptions the generating function $g$ for the branching weights,
\beq{geng}
g(z) = \sum_{n=1}^\infty w_n z^{n-1}\,,
\eeq
is strictly increasing and strictly convex on the interval $[0,\rho )$,
where $\rho$ is the radius of convergence for the series (\ref{geng}),
which we assume is positive.

It is well known, see e.g.\ \cite{book}, that the generating function
for the finite volume partition functions,
\beq{genZ}
Z(\zeta ) = \sum_{N=1}^\infty Z_N \zeta^N\,,
\eeq
satisfies the equation
\beq{fixZ}
Z(\zeta ) = \zeta g(Z(\zeta ))\,.
\eeq
The analytic function $Z(\zeta )$ which vanishes at
$\zeta=0$ is uniquely determined by \rf{fixZ}.  Letting $\zeta_0$ 
denote the radius
of convergence of the series (\ref{genZ}), the limit
\beq{x2}
Z_0 = \lim_{\zeta\uparrow \zeta_0}Z(\zeta )
\eeq
is finite and $\leq \rho$. In the following we consider the case where
\beq{genass}
Z_0<\rho\,.
\eeq
This is the condition on the branching weights which singles out the 
generic ensembles of infinite
trees to be defined below.  In particular, 
all sets of branching weights with infinite $\rho$ define a generic 
ensemble.

Assuming \rf{genass} the value of $Z_0$  is determined 
as the unique solution to the equation
\beq{eqcrit1}
Z_0g'(Z_0) = g(Z_0)
\eeq
and $\zeta_0$ can then be found from  \rf{fixZ}. 
Taylor expanding $g$ around $Z_0$ in
(\ref{fixZ}) and using (\ref{eqcrit1}) yields the well known singular
behavior of $Z$ at $\zeta_0$,
\beq{singZ}
Z(\zeta ) = Z_0 - \sqrt{\frac{2g(Z_0)}{\zeta _0g''(Z_0)}}\sqrt{\zeta_0-\zeta}+
O(\zeta _0-\zeta)\,.
\eeq
We shall need the following result on the asymptotic behavior of
$Z_N$, the proof of which can be found in \cite{flajosedg}, Sections VI.5
and VII.2.

\medskip
\noindent{\bf Lemma 1}\emph{
Under the stated assumptions on the branching weights and assuming
(\ref{genass}) the asymptotic behaviour of $Z_N$ is given by
\beq{u11}
Z_N= \sqrt{\frac{g(Z_0)}{2\pi g''(Z_0)}}N^{-\frac
32}\zeta _0^{-N}(1+O(N^{-1}))\,, 
\eeq
provided the integers $n$ for which $w_{n+1}\neq 0$ have no 
common divisor $>1$. 
Otherwise, if $d\geq 2$ denotes their largest common divisor, we have
\beq{u12}
Z_N= d\sqrt{\frac{g(Z_0)}{2\pi g''(Z_0)}}N^{-\frac
32}\zeta_0^{-N}(1+O(N^{-1}))\,, 
\eeq
if $N\equiv 1$ mod $d$, and $Z_N =0$ otherwise.}

\medskip

We define the probability distribution $\nu_N$ on $\Gamma _N$ by
\beq{p1}
\nu_N(\tau) = Z_N^{-1}\prod_{i\in\tau\setminus
r}w_{\sigma_i}\,,\quad\tau\in\Gamma_N\,,
\eeq
provided $Z_N\neq 0$.
Using Lemma 1 the existence of a limiting probability
measure $\nu$ on $\Gamma$ can be established by a minor modification of the
arguments in \cite{bergfinnur}, where the existence was proven for the
{\it uniform trees} corresponding to the weight factors $w_n=1$ for all
$n$. We state the result in the following theorem, providing an outline of
the proof in Appendix A.

\medskip

\noindent{\bf Theorem 2}\emph{
Viewing $\nu_N$ as a probability measure on $\Gamma$ we have, under
the same assumptions as in Lemma 1, that 
\beq{p2}
\nu_N \to \nu\quad \text{as}\quad N\to\infty\,,
\eeq
where $\nu$ is a probability measure on $\Gamma$ concentrated on the
subset $\Gamma_\infty$.} 

\medskip

\noindent
Here the limit should be understood in the weak sense, meaning that
\beq{p3}
\int_\Gamma f\,d\nu_N \quad \to\quad \int_\Gamma f\,d\nu
\eeq
for all bounded functions $f$ on $\Gamma$ which are continuous in the
topology defined by the metric $d_\Gamma$.  Moreover, $N$ is restricted to
values such that $Z_N\neq 0$.  We call the
ensembles $(\Gamma,\nu)$ {\it generic ensembles of
infinite trees}, referring back to the genericity assumption (\ref{genass}).
The expectation w.r.t.\ $\nu$ will be denoted $\langle\cdot\rangle _{\nu}$.

There is a simple description of
$\nu$ which is analogous to the description provided in
\cite{bergfinnur} for the measure on uniform trees. 
Given an infinite tree $\tau$ a {\it spine} is an infinite 
linear chain (non-backtracking path) in $\tau$ starting at the root. 
The result that $\nu$ 
is concentrated on the subset of trees with a single spine
is of crucial importance; it enables us to assume
that all infinite trees
have a unique spine.  
We denote the vertices on the spine
by $s_0=r, s_1, s_2,\dots$, ordered by increasing distance from the root.
To each $s_n ,\,n\geq 1$, are attached a finite number of {\it
branches}, i.e.\ finite trees in $\Gamma$, by identifying their roots
with $s_n$. If $s_n$ is of order $\sigma$ there are $\sigma-2$
branches attached to the spine at $s_n$. 
Let $T'_1,\dots, T'_k$ denote those to the
left (relative to the direction from $r$ along the spine) and
$T''_1,\dots, T''_\ell$ those to the right, ordered clockwise around
$s_n$, see Fig.\ 1.  
\begin{figure}[h!]
\begin{center}
\psfrag{r0}{$r$}
\psfrag{s1}{$s_1$}
\psfrag{s2}{$s_2$}
\psfrag{s3}{$s_3$}
\psfrag{s4}{$s_4$}
\includegraphics{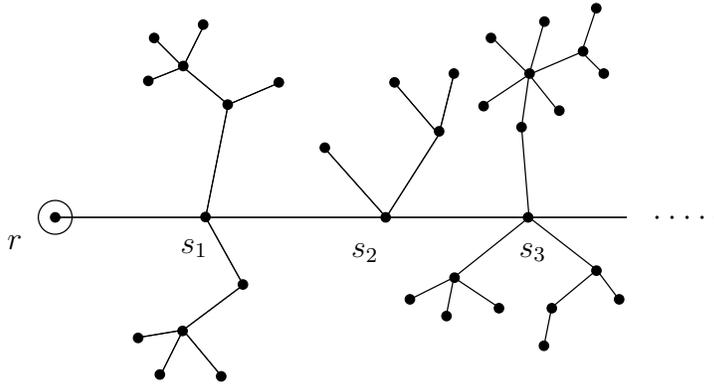}
\caption{The first few vertices on the spine of a tree and the finite branches
attached.}
\label{fig1}
\end{center}
\end{figure}
It follows from \rf{convVI} in Appendix A that
the probability that $s_n$ has $k\geq 0$ 
left branches and $\ell\geq 0$ right branches is
\beq{branchprob}
\varphi(k,\ell) = \zeta_0w_{k+\ell+2}Z_0^{k+\ell}\,,
\eeq
for all vertices $s_n$ on the spine.  
Note that this probability only depends on
$k+\ell =\sigma_{s_n} -2$.  Finally, the individual branches at
$s_n$ are independently distributed, and for each of them the
probability that a particular finite tree $T$ occurs is
\beq{mucrit}
\mu(T)= Z_0^{-1}\zeta_0^{|T|}\prod_{i\in T\setminus r}w_{\sigma_i}\,.
\eeq
We note that 
$\mu$ is the grand canonical distribution with
fugacity $\zeta_0$.

\section{Generic trees and Galton-Watson processes} 
The probabilities  $\mu (T)$ defined in 
\rf{mucrit} can be viewed 
as arising from a critical {\it
Galton-Watson process} as we show in Lemma 2 below. 
A Galton-Watson (GW) process is 
specified by a sequence $p_n,\,n= 0,1,2,\dots$, of non-negative numbers which
are called {\it offspring
probabilities} and satisfy 
\beq{prob1}
\sum_{n=0}^\infty p_n = 1\,.
\eeq
We say that the process is {\it critical} if the mean number of offspring is 
1, i.e.,         
\beq{GWcrit}         
\sum_{n=1}^\infty np_n = 1\,.         
\eeq 
A critical GW process gives rise to a 
probability distribution $\pi$ on the subset
of finite trees $T$ in $\Gamma$, see e.g.\ \cite{harris}, by 
\beq{GWprob}
\pi (T) = \prod_{i\in T\setminus r}p_{\sigma_i-1}\,.
\eeq

\medskip

\noindent{\bf Lemma 2}\emph{
Suppose the branching weights $w_n$ correspond to a generic ensemble of
infinite trees.
Consider the probability distribution $\mu$ \rf{mucrit} on
the set of finite trees in $\Gamma$.
Then $\mu$ corresponds to a critical 
Galton-Watson process with offspring probabilities 
\beq{GWprobw}
p_n = \zeta _0 w_{n+1}Z_0^{n-1}\,.
\eeq  }

\medskip   

\noindent{\it Proof.} With $p_n$ given by (\ref{GWprobw}) we get
\beq{x3}
\sum_{n=0}^\infty p_n = \zeta _0 \sum_{n=0}^\infty w_{n+1}Z_0^{n-1}\; =\; 
\zeta _0Z_0^{-1}g(Z_0)\;=1\,,
\eeq
where the last equality follows from (\ref{fixZ}).  Furthermore, by \rf{GWprob},
\beq{x4}
\pi (T)\; = \;\zeta _0^{|T|}\prod_{i\in T\setminus r}
w_{\sigma_i}Z_0^{\sigma_i-2}
\;= \;Z_0^{-1} \zeta _0^{|T|}\prod_{i\in T\setminus r}w_{\sigma_i}\,,
\eeq
since
\beq{x5}
\sum_{i\in T\setminus r}(\sigma_i-2)\;=\; -1\,,
\eeq
for a tree $T$ with a root of order $1$. This proves that
$\pi (T)=\mu (T)$.  Finally,
\beq{x6}
\sum_{n=1}^\infty np_n\; =\; \zeta _0 \sum_{n=1}^\infty nw_{n+1}Z_0^{n-1}\;
=\; \zeta _0g'(Z_0)\,,
\eeq
which proves, in view of \rf{fixZ} and \rf{eqcrit1},  that 
the process is critical.
\hfill $\square$

\medskip  

In the following we let $f$ denote the generating function for the
offspring probabilities given by (\ref{GWprobw}),
\beq{genf}
f(z)= \sum_{n=0}^\infty p_nz^n \; =\; \zeta_0\sum_{n=1}^\infty
w_nZ_0^{n-2}z^{n-1}\; =\; \zeta_0 Z_0^{-1}g(Z_0z)\,.
\eeq
Equations\ (\ref{prob1}) and (\ref{GWcrit}), or equivalently (\ref{fixZ}) for
$\zeta =\zeta_0$ and (\ref{eqcrit1}), may then be written 
\beq{fcrit}
f(1) = 1\quad\text{and}\quad f'(1)=1\,,
\eeq
respectively. Moreover, the genericity assumption (\ref{genass}) is
equivalent to {\it assuming $f$ to be analytic in a neighbourhood of
the unit disc.}

If $T$ is a finite tree, let $h(T)$ denote its {\it height},
i.e.\ the maximal distance from the root $r$ to a vertex in $T$.
As a consequence of Lemma 2 we note the following
result.

\medskip

\noindent{\bf Lemma 3}\emph{ 
Let $\mu$ be the measure on finite trees 
given by (\ref{mucrit}) and let $\langle\cdot\rangle_\mu$
denote the expectation w.r.t.\ $\mu$. Then 
\beq{heightdistr}
\mu(\{T\in\Gamma\mid \; h(T)> R\}) = \frac{2}{f''(1)R} + O(R^{-2})
\eeq 
for $R$ large.
Moreover,
\beq{ballexp}
\langle |B_R|\rangle_\mu = R\;.
\eeq
}

\medskip

\noindent{\it Proof.}
Both properties are standard consequences of the fact that $\mu$ is a 
critical GW process, see e.g.\ \cite{harris}, Sections I.5 and
I.10. \hfill $\square$  

\medskip

For $\tau\in\Gamma$ let $D_R(\tau )$ denote the number of vertices in
$\tau$ at distance $R$ from the root.
The relation between $\nu$ and $\mu$ described above gives
rise to the following useful result.

\medskip

\noindent{\bf Lemma 4}\emph{
Let $(\Gamma,\nu)$ be a generic ensemble of infinite trees with
corresponding critical Galton-Watson measure $\mu$.
For any bounded function $u$ on $\Gamma$ such that
$u(\tau)=u(B_R(\tau)),\,\tau\in\Gamma$, i.e.\
 $u(\tau)$ depends only on
the part of $\tau$ contained in the ball of radius $R$ around the
root, we have}
\beq{munu}
\int_\Gamma u(\tau)d\nu(\tau) = \int_\Gamma u(T)D_R(T)d\mu(T)\,.
\eeq

\medskip 

\noindent{\it Proof.}
Let $\tau\in \Gamma$ and let $T$ denote the
finite subtree of $\tau$ 
consisting of the first $R+1$ vertices $s_0,s_1,s_2,\ldots
,s_R$  on the spine 
together with all the branches attached to these vertices.  Then
$u(\tau )=u(T)$.  Let  
$\Gamma(R)$ denote the set of 
all rooted finite planar trees with a marked vertex
$s$, of order one, at a distance $R$ from the root.
Then we can write the left hand side of
(\ref{munu}) as a sum over
$\Gamma(R)$ by performing the integration over the branches attached
to the vertex $s_R$ on the spine including the infinite one containing
the part of the spine beyond $s_R$, which is distributed
according to $\nu$. The result is 
\beq{munu1}
\sum_{T\in\Gamma(R)}
u(T)\prod_{i\in T\setminus\{r,s_R\}}p_{\sigma_i-1}\,. 
\eeq

On the other hand, the right hand side of \rf{munu} 
is a sum over finite rooted
trees, where only trees of height at least $R$ contribute because of the
factor $D_R(T)$. This sum can be replaced by a sum over finite
rooted trees with one marked vertex at distance $R$ from the root upon
deleting the factor $D_R(T)$. Summing over the branches containing
the descendants of the marked vertex again yields the sum
(\ref{munu1}) thereby establishing the lemma. \hfill $\square$

\medskip

The integration formula \rf{munu} has the following application which
will be needed in the next section.

\medskip

\noindent{\bf Lemma 5}\emph{
Let $(\Gamma,\nu)$ be a generic ensemble of infinite 
trees. Then there exists a constant $c>0$ such that
\beq{z1}
\langle\, |B_R|^{-1}\,\rangle_\nu\; \leq\;  c\,R^{-2}
\eeq
for all $R\geq 1$.}

\medskip

\noindent{\it Proof.}
Using the function
\beq{z2}
u(\tau)= \begin{cases} D_R(\tau)^{-1}\;&\text{if $D_R(\tau)\neq
0$}\\
\;\;0\;&\text{if $D_R(\tau)=0$}\,,
\end{cases} 
\eeq
in Lemma 4 we get from (\ref{heightdistr}) that 
\beq{z3}
\langle\, D_R^{-1}\,\rangle_\nu\; \leq\; \frac{c'}{R}
\eeq
for $R\geq 1$, where $c'$ is a positive constant. From this we conclude
\begin{eqnarray}
\langle |B_R(\tau)|^{-1}\rangle_\nu &=& \left\langle
\frac{1}{D_1+\dots+D_R}\right\rangle_\nu\nonumber\\
&\leq& R^{-1} \left\langle (D_1D_2\dots D_R)^{-\frac 1R}\right\rangle_\nu
\nonumber\\
&\leq& R^{-1}\prod_{i=1}^R\left\langle\, D_i^{-1}\,\right\rangle^{\frac
1R}_\nu
\nonumber\\
&\leq& c'R^{-1}(\,R!\,)^{-\frac 1R}\nonumber\\
&\leq& cR^{-2}\,. 
\end{eqnarray} \hfill$\square$
 
\medskip

\noindent{\bf Remark.}
Lemma 5 can easily be strengthened to
\beq{z5}
c_1R^{-2}\leq \langle\, |B_R|^{-1}\,\rangle_\nu \leq c_2 R^{-2}\,,
\eeq
where $c_1$ and $c_2$ are positive constants.
Indeed, by Jensen's inequality, we have
\beq{z6}
 \langle\, |B_R|^{-1}\,\rangle_\nu\;\geq\; (\langle\,
|B_R|\,\rangle_\nu)^{-1}\,, 
\eeq
and from standard arguments using generating functions one has
\beq{z7}
\langle\, |B_R|\,\rangle_\nu = \frac 12 f''(1)R(R-1) +R \,.
\eeq
Equation \rf{z7} also shows that the Hausdorff dimension, see \rf{haus},
of the ensemble
$(\Gamma,\nu)$ is $2$. 

\medskip

\section{Proof of Theorem 1}

For $\tau\in\Gamma$ let $\omega $ be a random walk on $\tau$
starting at the root at time $0$.
Let $\omega (t)$ denote the vertex where $\omega $
is located after $t$ steps, $t\leq |\omega |$.
The generating function
for return probabilities of simple random walk on $\tau$,
$Q_\tau(x)$, is given by
\beq{Qdef}
Q_\tau(x) = \sum_{\omega:r\to
r}(1-x)^{\oh |\omega|}\prod_{t=1}^{|\omega |-1}\sigma_{\omega (t)}^{-1},
~~~0<x\leq 1,
\eeq 
where the sum is over all walks in $\tau$ starting and ending at the
root $r$, including the trivial walk consisting of $r$ alone which
contributes $1$ to $Q_\tau (x)$.
The generating function for first return
probabilities $P_\tau (x)$ is given by \rf{Qdef} except that
the sum excludes the trivial walk and
is restricted to walks that do not visit $r$ in between the
initial and final position.
The functions $Q_\tau (x)$ and $P_\tau (x)$ are related by the identity
\beq{PQ}          
Q_\tau(x) = \frac{1}{1-P_\tau(x)}\,.          
\eeq
This equation implies that $P_\tau (x)\leq 1$.
There is a more general recurrence relation which we will use repeatedly in
the rest of the paper.  Let $\tau\in\Gamma$ and $v$ the vertex next
to the root.  Let $\tau_1,\tau_2,\ldots ,\tau _k\in\Gamma$ 
be the subtrees of $\tau$ 
meeting the link $(r,v)$ at $v$.  
By decomposing walks from the root
and back into a sequence of excursions into the different branches 
$\tau_i$
of the tree $\tau$ we find 
\beq{PPP}
P_\tau (x)={1-x\over k+1-\sum_{i=1}^kP_{\tau _i}(x)}.
\eeq

For the ensemble  $(\Gamma,\nu)$ we set
\beq{z9}
Q(x) = \langle Q_\tau(x)\rangle_\nu
\eeq
and define the critical exponent $\alpha$ associated with $Q(x)$ by
\beq{defalpha}
Q(x) \sim x^{-\alpha}
\eeq
as $x\to 0$.   By $a(x)\sim x^\beta$ for $x\to 0$ 
we mean that 
for any $\epsilon >0$ there are constants $c_1$ and $c_2$, which may depend
on $\epsilon$, 
such that for $x$ small enough 
\beq{z11}
c_1x^{\beta +\epsilon}\leq a(x)\leq c_2x^{\beta -\epsilon}.
\eeq
This gives a precise meaning to the definition of the spectral dimension 
\rf{specdim0} and,
assuming it exists, 
$d_s$ is related to $\alpha$ by
\beq{specdim}
d_s = 2-2\alpha\,.
\eeq
Below we will prove a stronger result than \rf{defalpha} and
show that there exist positive constants
$\underline{c}$ and $\overline{c}$ such that for $x$ small enough
\beq{x11}
\underline{c}\,x^{-1/3}\leq Q(x)\leq \overline{c}\,x^{-1/3}.
\eeq

\subsection{Lower bound on $Q(x)$}

In order to establish the relevant lower bound we need two 
preliminary lemmas.  

\medskip

\noindent{\bf Lemma 6}\emph{
For all finite trees $T\in\Gamma$ and $0<x\leq 1$ we have}
\beq{t66}
P_T (x) \geq 1-|T|x\,.
\eeq

\smallskip

\noindent{\it Proof.}  Let $T_1$, $T_2, \ldots , T_{n-1}$ be the trees attached
to the vertex $v$ of $T$ next to the root.  Then from \rf{PPP} we obtain
\beq{ccc}
P_T (x)={1-x\over n-\sum_{i=1}^{n-1}P_{T_i}(x)}.
\eeq
The first return generating function for the tree consisting of a single
link is $1-x$.  The lemma follows by induction on $|T|$.
\hfill$\square$
 
\medskip

\noindent{\bf Lemma 7}\emph{
Let $\tau\in \Gamma$ be a tree with one infinite spine. For all $L\geq
1$ and $0<x\leq 1$ we have
\beq{x13}
P_\tau(x)\;\geq\; 1-\frac 1L - Lx -{\sum_{T\subset\tau}}^L(1-P_T(x))\,,
\eeq
where  $\sum_{T\subset\tau}^L$ denotes the sum over all (finite)
branches $T$ of $\tau$ attached to vertices on the spine at distance
$\leq L$ from the root.}

\medskip

\noindent{\it Proof.}
We give an inductive argument proving the stronger inequality 
\beq{x14}
P_\tau^L(x)\;\geq\; 1-\frac 1L - Lx -{\sum_{T\subset\tau}}^L(1-P_T(x))\,,
\eeq
where $P_\tau^L(x)$ denotes the contribution to $P_\tau(x)$ from walks
$\omega$ that do not visit the vertex $s_{L+1}$ on the spine,
i.e.\ walks constrained to the the first $L$ vertices on the spine
after the root and the branches attached to them.

The inequality holds for $L=1$, since
the right hand side of \rf{x14} is non-positive in this case.
For $L\geq 2$ we have from \rf{PPP} 
\beq{rec}
P_\tau^L(x) = \frac{1-x}{n -
P_{\tau_1}^{L-1}(x)-\sum_{k=1}^{n-2}P_{T_k}(x)}\,,
\eeq
where $n=\sigma_{s_1}$ is the order in $\tau$ of the vertex $s_1$ and
$T_1,\dots, T_{n-2}$ denote the (finite) branches attached to $s_1$,
while $\tau_1$ is the infinite branch attached to $s_1$, i.e.\ the
subtree with root $s_1$ and containing $s_2$ and all its
descendants. 

Using the induction hypothesis 
\beq{ineqP}
P_{\tau_1}^{L-1}(x)\;\geq\; 1-\frac 1{L-1} - (L-1)x -
{\sum_{T\subset\tau_1}}^{L-1}(1-P_T(x))\,,
\eeq
we get from (\ref{rec})
\begin{eqnarray}
P_\tau^L(x) &=& \frac{1-x}{1 +
(1-P_{\tau_1}^{L-1}(x))+\sum_{k=1}^{n-2}(1-P_{T_k}(x))}\nonumber\\
&\geq& 
\frac{1-x}{1 +\frac 1{L-1}+ (L-1)x +\sum_{T\subset\tau}^L(1-P_{T}(x))}
\nonumber\\
&\geq& 
\frac{L-1}{L}\,\frac{1-x}{1+(L-1)x +\sum_{T\subset\tau}^L(1-P_{T}(x))}
\nonumber\\
&\geq& 
(1-\frac 1L)(1-x)\left(1-(L-1)x -{\sum_{T\subset\tau}}^L(1-P_{T}(x))\right)
\nonumber\\
&\geq& 
1-\frac 1L - Lx -{\sum_{T\subset\tau}}^L(1-P_{T}(x))\,.\label{longarray}
\end{eqnarray}
Here we have assumed for the last inequality that the final expression
is positive. Otherwise, the inequality (\ref{ineqP})
holds trivially. This proves the lemma. \hfill$\square$

\medskip

We are now ready to establish the desired lower bound \rf{x11} on $Q(x)$.
The argument combines Lemmas 3, 6 and 7 with Jensen's inequality.
Let $s$ be any vertex on the spine different from $r$. Given that
$s$ has $k$ left branches and $\ell$ right branches the probability
that a given branch has height $>R$ is given by
(\ref{heightdistr}). Hence the conditional probability $c_R$ that at
least one of the $k+\ell$ branches has height $>R$ fulfills
\beq{xy}
c_R\leq (k+\ell )\left({2\over f''(1)R}+O(R^{-2})\right).
\eeq
According to (\ref{branchprob}) the $\nu$-probability $q_R$ 
that at least
one branch at $s$ has height $>R$ then fulfills
\beq{z20}  
q_R\leq 
\left({2\over f''(1)R}+O(R^{-2})\right)
\sum_{k,\ell\geq 0}(k+\ell)\varphi(k,\ell)\;
=\; \frac{2}{R} +O(R^{-2})\,.
\eeq

By independence of the distribution of branches attached to different
vertices on the spine we conclude that the $\nu$-probability of the
event ${\cal A}_R$, that all
branches attached to the first $R$ vertices $s_1,\dots, s_R$ on the
spine have height $\leq R$, satisfies
\beq{p9}
\nu({\cal A}_R) = (1-q_R)^R 
\geq\exp \left(-2+O(R^{-1})\right).
\eeq
Denoting by $\langle\cdot \rangle_R$ the expectation 
w.r.t.\ $\nu$ conditioned on the event ${\cal A}_R$, we get by Lemmas 6
and 7 and Jensen's inequality that
\begin{eqnarray}\label{est1}
Q(x)&\geq & e^{-2 +O(R^{-1})}
\langle (1-P_\tau(x))^{-1}\rangle_R\nonumber\\
&\geq&  e^{-2 +O(R^{-1})}  \left\langle \left(\frac 1R + Rx + {\sum_{T\subset\tau}}^R
x|T|\right)^{-1}\right\rangle_R\nonumber\\ 
&\geq&  e^{-2 +O(R^{-1})}   \left(\frac 1R + Rx + x\left\langle{\sum_{T\subset\tau}}^R
|T|\right\rangle_R\right)^{-1}\,.
\end{eqnarray}

Let $B_R^i(\tau )$ denote the subgraph of $\tau $ which is spanned by all
vertices whose distance from the vertex $s_i$ is at most $R$ and 
which lie in the branches rooted at $s_i$.
Using again that the distributions of branches are identical at all vertices
on the spine and given by (\ref{branchprob}) and (\ref{mucrit}), we have
\begin{eqnarray}  
\left\langle{\sum_{T\subset\tau}}^R |T|\right\rangle_R
&=&\left\langle\sum_{i=1}^R |B_R^i(\tau )|\right\rangle_R \nonumber\\ 
&\leq& (1-q_R)^{-1}R\langle|B_R^1|\rangle_\nu\nonumber\\
&=& (1-q_R)^{-1}R\sum_{k,\ell\geq
0}\varphi(k,\ell)(k+\ell)\langle|B_R|\rangle_\mu\nonumber\\
&=& \frac{f''(1)}{1-q_R}R^2.
\end{eqnarray}
Inserting the last estimate into (\ref{est1}) and 
observing from \rf{z20} that $q_R\to 0$ as
$R\to \infty$ we deduce
\beq{p10}
Q(x)\geq
c' \left(\frac 1R + Rx + f''(1) x R^2\right)^{-1}\,,
\eeq
where $c'$ is a positive constant, for $R$ large enough. 
Finally, choosing $R=[x^{-1/3}]$ yields
\beq{p11}
Q(x)\;\geq\;\underline{c}\,x^{-1/3}
\eeq
for a suitable constant $\underline{c}>0$ and $x$ small enough, 
as claimed. \hfill$\square$

\subsection{Upper bound on $Q(x)$} 

We begin by establishing a monotonicity lemma which is a slight
generalization of the corresponding result in \cite{DJW}.

\bigskip

\noindent
{\bf Lemma 8}  {\it Let $\tau$ be a rooted 
tree, $\omega_v$ the shortest path on
$\tau$ from $r$ to the vertex $v$ and let $v_j$ be the $j$th vertex from $r$
along $\omega_v$, $j=1,2,\ldots ,|\omega _v|$.  Denote by
$\tau_{j1},\tau_{j2},\ldots ,\tau_{jK_j}$ the subtrees of $\tau$ 
attached to the vertex
$v_j$ which do not contain any link in $\omega_v$ and such that their roots
$v_j$ have order $1$.
 Then $P_\tau$ is an increasing function of
each $P_{\tau_{jk}}$.  In particular, if $\tau '$ is the tree obtained
by removing one of the $\tau_{jk}$ from $\tau$, then  
$P_{\tau^{'}}(x)\geq P_\tau (x)$. 
}

\bigskip

\noindent
{\it Proof.} 
Let $\tau_j$ denote the 
rooted tree with root $v_j$ obtained
from            
$\tau$ by amputating all the branches $\tau_{ik}$, $i=1,2,\ldots ,j$, 
and the links $(r,v_1),(v_1,v_2),\dots,(v_{j-1},v_j)$.   From \rf{PPP} we have
the recursion
\beq{recursion}
P_{\tau }(x)={1-x\over K_1+2-P_{\tau_{1}}(x)-\sum_{k=1}^{K_1}P_{\tau_{1k}}(x)}.
\eeq
We see that $P_{\tau }$ is an increasing function of $P_{\tau_{1}}$ and 
$P_{\tau_{11}}, P_{\tau_{12}},\ldots ,P_{\tau_{1K_1}}$.  The lemma
follows by induction.
\hfill$\square$
\bigskip

The upper bound will be obtained from the above lemma
and some elementary estimates.    Let $\tau\in
\Gamma$. Define $p_\tau (t;v)$ to be the probability that a random walk
which starts at the root at time $0$ is at the vertex $v\in\tau$ 
after $t$ steps.  That is,
\beq{x555}
p_\tau (t;v)=\sum_{\omega :r\to v}\prod_{s=1}^{t-1}\sigma_{\omega (s)}^{-1},
\eeq
where the sum is over all walks of length $t$ from $r$ to $v$.
 Define the corresponding generating function
\beq{z22}
Q_\tau (x;v)=\sum_{t=0}^\infty p_\tau (t;v)(1-x)^{t/2},~~~0<x\leq 1.
\eeq
Note that $Q_\tau (x)=Q_\tau (x;r)$.

Summing \rf{z22} over a ball of radius $R$ centred on the root gives
\beq{z23}
\sum_{v\in B_R (\tau )}Q_\tau (x;v)\leq\sum_{t=0}^\infty 
(1-x)^{t/2}\leq {2\over x}.
\eeq
It follows that there is a vertex
$\bar{v}\in B_R(\tau )$ such that
\beq{z24}
Q_\tau (x;\bar{v})\leq {2\over x|B_R (\tau )|}.
\eeq  
If $\bar{v}\neq r$ 
we can split the random walk representation of $Q_\tau (x)$ into two parts:
walks that do not reach the vertex $\bar{v}$ and walks that do.  Let us
denote the first contribution by $Q^{(1)}_\tau (x)$ and the second one by 
$Q^{(2)}_\tau (x)$.  Let $L\geq 1$ 
denote the distance of $\bar{v}$ from the root.
Then by Lemma 8 we have
\beq{z25}
Q^{(1)}_\tau (x)\leq {1\over 1-R_L(x)},
\eeq
where $R_L(x)$ is the generating function for first return to the root of
walks on the integer half line which are restricted not to move beyond the
$(L-1)$st vertex.  This function can be calculated explicitly, see \cite{DJW}
Appendix B,  with the result
\beq{z26}
R_L(x)= (1-x){(1+\sqrt{x})^{L-1}-(1-\sqrt{x})^{L-1}\over
(1+\sqrt{x})^{L}-(1-\sqrt{x})^{L}}.
\eeq
It is straightforward to show that
\beq{straight}
{1\over 1-R_L(x)}\leq L
\eeq
for $0<x\leq 1$.  Hence,
\beq{z27}
Q^{(1)}_\tau (x)\leq L\leq R.
\eeq

If $v,v'$ are two different vertices in $\tau$ we
define $G_\tau (x;v,v')$ by \rf{Qdef} with the walks restricted
to start at the vertex $v$, end at $v'$ and not visit
$v$ again.
Any walk $\omega$ that contributes to $Q^{(2)}_\tau (x)$ can be split uniquely
into two parts, an arbitrary walk $\omega_1$ 
from the root to $\bar{v}$ and a walk $\omega_2$ from $\bar{v}$ back to the
root which does
not revisit $\bar{v}$.  Hence,
\beq{z28}
Q^{(2)}_\tau (x) = \sigma_{\bar{v}}^{-1} 
Q_\tau (x;\bar{v})G_\tau (x;\bar{v},r).
\eeq
Let $v$ be any vertex in $\tau\in\Gamma$ different from $r$.  Let $\omega_v$
be the shortest path from $r$ to $v$ and $v_0=r,v_1,\ldots ,v_n$ its
vertices, $n=|\omega _v|$.
Then we have the representation
\beq{n11}
G_\tau(x;v) = \sigma_{v}(1-x)^{-|\omega_v|/2}\prod_{k=0}^{n-1}P_{\tau_k}(x)
\eeq  
which is easily obtained by decomposing a walk $\omega$ from $r$ to $v$
into $n$ 
walks $\omega^k,\,k=0,1,\dots, n-1$, such that $\omega^k$ starts
at $v_k$ and ends at $v_{k+1}$ and avoids $v_k$ after leaving it, see
\cite{DJW} Section 2.2 for details.
In \rf{n11} the trees $\tau_k$ are defined as in the proof of Lemma 8.
Applying \rf{n11}, with the roles of $r$ and $v$ interchanged, 
and Lemma 8 we see that
\beq{y777}
G_\tau (x;v,r)\leq G_{\tilde{\tau}}(x;v,r)
\eeq
for any vertex $v\in\tau$ where $\tilde{\tau}$ is the chain of links forming
the shortest path from $v$ to $r$.  It is straightforward to compute 
$G_{\tilde{\tau}}(x;v,r)$, see \cite{DJW}, with the result
\beq{y778}
G_{\tilde{\tau}}(x;v,r)={2(1-x)^{L/2}\over (1+\sqrt{x})^L+(1-\sqrt{x})^L}
\leq 1.
\eeq
We conclude from \rf{z24}, \rf{z27} and \rf{z28} 
that
\beq{z29}
Q_\tau (x)\leq R+{2\over x|B_R (\tau )|}.
\eeq
Obviously, this inequality also holds if $\bar{v}=r$.
Taking the expectation of the above inequality 
with respect to the measure $\nu$ and using Lemma 5 yields
\beq{z30}
Q(x)\leq R+{c\over xR^2}
\eeq
with $c$ a positive constant.
This establishes the desired upper bound \rf{x11} on $Q(x)$ by choosing 
$R=[x^{-1/3}]$ and completes the proof of Theorem 1.
\hfill$\square$

\medskip
\noindent
{\bf Remark.}
The results about the spectral dimension of generic trees generalize to the
case when the root $r$ has a fixed order $m\geq 1$.  Let $\Gamma^{(m)}$
denote the set of planar trees with a distinguished {\it root link} $(r,r')$
where $r$ has order $m$.  We define the partition functions $Z_N^{(m)}$,
$N\geq m$, by the right hand side of \rf{x1} with $\Gamma_N$ replaced by
$\Gamma_N^{(m)}=\{ \tau\in\Gamma_N : |\tau |=N\}$.  The corresponding
generating function $Z^{(m)}(\zeta )$ is then given by $Z^{(m)}(\zeta )
=Z(\zeta )^m$
where $Z(\zeta )$ is as in \rf{genZ}.  This relation implies an immediate
generalization of Lemma 1 and also the existence of a probability measure
$\nu^{(m)}$ supported on the subset of $\Gamma^{(m)}$ consisting of trees
with one infinite spine originating at $r$.  This measure can be
characterized in the same way as $\nu$.  In particular, the (finite)
branches have the same probability distribution as in the $m=1$ case and the
$m$ branches (including the infinite one) originating at $r$ have equal
probabilities of being infinite.  Using this observation the arguments of
Section 4 can be carried through with minor modifications to yield the same
upper and lower bounds \rf{x11} for the generating function $Q^{(m)}(x)$ for
return probabilities to the root $r$.  Similar remarks apply to the case
where the root is subject to the same degree distribution as the other
vertices of the tree.

\section{The mass exponent of generic random trees}

Let us consider an infinite tree $\tau$ 
with a single spine.  We label the vertices
on the spine as before $s_0=r,
s_1,s_2,\ldots $.  For convenience let us introduce
the notation $Q_\tau (x;s_n)=Q_\tau (x;n)$, $n\in\bbN$.
In the same way we write $G(x;s_k,s_n)=G(x;k,n)$ and $G(x;s_n)=
G(x;r,s_n)=G(x;n)$.
The purpose of this section is to determine the critical behaviour of
the mass $m(x)$, defined as the exponential decay rate of the {\it two-point
function}  $Q(x;n)$, defined by
\beq{tpf}
Q(x;n) = \langle Q_\tau(x;n)\rangle_\nu\,,
\eeq
as a function of $n$, i.e.,
\beq{massdef}
m(x) = -\lim_{n\to\infty}\frac{\log Q(x;n)}{n}\,.
\eeq
 We show below that the limit above exists and is positive for
$0<x<1$ and tends to zero as $x\to 0$. The critical exponent $\nu$ 
of the mass is defined by
\beq{w1}
m(x) \sim x^\nu
\eeq
as $x\to 0$, provided it exists.  The methods of this paper do not suffice
to show the existence of $\nu$. We prove that if $\nu$ exists then 
\beq{w2}
\nu = \alpha = 1/3\,,
\eeq
where $\alpha$ is defined by (\ref{defalpha}).

First, let us establish existence of the mass.

\medskip

\noindent{\bf Lemma 9}\emph{
For a generic ensemble $(\Gamma,\nu)$ of infinite trees the mass
$m(x)$ is well defined by (\ref{massdef}). It is a non-decreasing
function of $x$ and fulfills
\beq{expmass}
x^{\frac 12}\,e^{-m(x)n}\leq Q(x;n)\leq C\tilde{C}^{-1}\,x^{-1}\,e^{-m(x)n}
\eeq
for $n\geq 1$,  where $C=f''(1)+2$ and $\tilde{C}=\sum_{n,m\geq 0}{\varphi
(n,m)\over n+m+2}$.
}

\medskip

\noindent{\it Proof.}  
Note first that, if $\tau$ is a rooted tree with a single spine,
$Q_\tau(x;n)$ can be written as 
\beq{w5}
Q_\tau(x;n) = Q_\tau(x)G_\tau(x;n)\,.
\eeq  
Similarly, for $i\geq 1$,
\beq{w6}
G_\tau(x;i) =  R_\tau(x;i)G^0_\tau(x;i)\,,
\eeq
where $G^0_\tau(x;i)$ is defined by the same formula (\ref{Qdef}) as
$Q_\tau(x)$ but with the walks $\omega$ going from the root to 
$s_i$ and restricted to visit
neither the root nor the endpoint $s_i$  except at their start and
end.  The function $R_\tau(x;i)$ is defined by the same formula as
$Q_\tau(x)$ except that the walks $\omega$ start and end at $s_i$ and
avoid the root.  In the definition of $R_\tau(x;i)$ we also include an extra
factor $\sigma_{s_i}^{-1}$ associated with the initial step.
It is easy to see that 
\beq{w7}
Q_\tau(x)\leq x^{-\oh}~~\mbox{\rm and}~~R_\tau(x;i)\,\leq\,\sigma_{s_i} 
x^{-\frac 12}\,
\eeq
using Lemma 8.
It follows that
\beq{w8}
G_\tau(x;i)\leq Q_\tau(x;i) \leq x^{-\frac 12} G_\tau(x;i)\,,
\eeq
and 
\beq{w9}
G^0_\tau(x;i)\leq G_\tau(x;i) \leq  \sigma_{s_i}x^{-\frac 12} G_\tau^0(x;i)\,.
\eeq
Taking averages w.r.t.\ $\nu$ in the above inequalities yields 
\beq{FG}
G(x;i)\leq Q(x;i) \leq x^{-\frac 12} G(x;i)
\eeq
and
\beq{GG0}
G^0(x;i) \leq G(x;i) \leq C\,x^{-\frac 12}\,G^0(x;i)\,,
\eeq
where  $G(x;i)$ and $G^0(x;i)$ denote 
$\langle G_\tau (x;i)\rangle_\nu$ and $\langle G^0_\tau (x;i)\rangle_\nu$
respectively, and the
constant $C$ is given by
\beq{w11}
C = \sum_{n,m=0}^\infty (n+m+2)\,\varphi(n,m) = f''(1)+2.
\eeq
By similar arguments, decomposing walks and the tree $\tau$ suitably,
we have, for $i,j\geq 1$, 
\beq{w12}
G_\tau^0(x;i) G_{\tau_i}^0(x;j) \leq \sigma_{s_i}G_\tau^0(x;i+j)\leq 
\sigma_{s_i}\, x^{-\frac 12}\, G_\tau^0(x;i) G_{\tau_i}^0(x;j)\,,
\eeq
where $\tau_i$ is defined as in the proof of Lemma 8 with $\omega_v$ the
path along the spine from $r$ to $s_i$.
Averaging w.r.t.\ $\nu$ then yields 
\beq{w13}
\tilde{C}G^0(x;i) G^0(x;j) \leq G^0(x;i+j)\leq 
x^{-\frac 12}\, G^0(x;i) G^0(x;j)\,,
\eeq
since $G_\tau^0(x;i)$ and $G_{\tau_i}^0(x;j)$ are independent random
variables.  

It follows that $-\log (\tilde{C}G^0(x;i))$ 
and  $\log (x^{-\frac 12} G^0(x;i))$
are subadditive  functions of $i$. Hence,
\beq{w16}
-\,\lim_{i\to\infty}\frac{\log G^0(x;i)}{i} =  -\,\sup_{i\geq
1}\frac{\log (\tilde{C}G^0(x;i))}{i} = -\, \inf_{i\geq
1}\frac{\log (\,x^{-\frac 12}\, G^0(x;i))}{i}\,.
\eeq
In view of (\ref{FG}) and (\ref{GG0}) this proves that the mass exists
and fullfills (\ref{expmass}). Since $G^0(x;i)$ (as well as $Q(x;i)$)
clearly is a decreasing function of $x$ it follows that $m(x)$ is
non-decreasing. \hfill$\square$

\smallskip


\noindent{\bf Theorem 3}\emph{ 
There exist positive constants
$c_1$ and $c_2$ such that
\beq{mest}
c_1{x^{1/3}\over |\log x|^{2/3}}\,\leq\, m(x)\,\leq\, 
c_2\, x^{1/3}|\log x|
\eeq
for $x$ sufficiently small.}

\smallskip

\noindent{\it Proof.}
Let $s_n$ be a vertex on the spine.  Then by \rf{z25}, \rf{z28} and 
\rf{y778} we obtain
\beq{uu88}
Q_\tau (x)\leq {1\over 1-R_n(x)} +\sigma_{s_n}^{-1}Q_\tau (x;n),
\eeq
with $R_n(x)$ as in \rf{z26}.  Taking the expectation value we get
\beq{uu89}  
Q(x)\leq n+Q(x;n).
\eeq
Now choose $n=[|\log x| m(x)^{-1}]+1$
in \rf{uu89}.
In view of (\ref{expmass}) we then have 
\beq{e17}
Q(x;n) \leq C\tilde{C}^{-1}\,.
\eeq
Using \rf{p11}  and \rf{uu89} 
we find that as $x\to 0$   
\beq{e18}
\underline{c}\,x^{-1/3}\leq |\log x|m(x)^{-1}+1+C\tilde{C}^{-1}\,,
\eeq
proving the claimed upper bound. 

To obtain the lower bound we make use of the representation 
\rf{n11}.
By \rf{w9} it follows that
\beq{Gest1}
\tilde{C}G^0(x;n)\, \leq\ (1-x)^{-n/2}\prod_{k=0}^{n-1}\,\langle
\,P_{\tau_k}(x)^n\,\rangle_\nu^{1/n}\, =\, (1-x)^{-n/2}\langle\,
P_{\tau}(x)^n\,\rangle_\nu\,,
\eeq
where we have used 
\beq{n12}
\langle\, P_{\tau_k}(x)^n\,\rangle_\nu\, =\, \langle\, 
P_{\tau}(x)^n\,\rangle_\nu
\eeq
which is 
a consequence of the characterisation of $\nu$ given in Section 2.

By (\ref{PQ}) and \rf{z29} we have 
\beq{Pest1}
P_\tau(x) \leq 1-\left(
\frac{2}{x|B_R(\tau)|} + R\right)^{-1}\,.
\eeq
Now, let the event ${\cal C}(\lambda,R)$, where $\lambda >0$
and $R\geq 1$,  be defined by
\beq{n14}
{\cal C}(\lambda,R) = \{\tau\in\Gamma\,|\; |B_R(\tau)| \geq \lambda R^2\}\,.
\eeq
It can be shown that 
\beq{Cset}
\nu(\Gamma\setminus{\cal C}(\lambda,R)) \leq e^{-c_0\lambda^{-1/2}}
\eeq
for some constant $c_0>0$ and $\lambda$ in an interval $(0,\lambda _0)$.
This is proven in a special case in
\cite{BarKum} Lemma 2.4, but the proof can be generalised in a
straightforward manner to arbitrary generic ensembles of infinite trees.
For the sake of completeness details are provided in Appendix B.

Setting $R=[x^{-1/3}|\log x|^{2\over 3}]$ and $\lambda = 
{c_0^2/4|\log x|^2}$ we get 
for $\tau\in{\cal C}(\lambda ,R)$ and small $x$,
\beq{n16}
P_\tau(x) \leq 1- {c''x^{1/3}\over |\log x|^{2/3}}
\eeq
where $c''$ is a positive constant.
Furthermore,
\beq{n17}
\nu(\Gamma\setminus{\cal C}(\lambda,R))\,\leq\, x^2\,.
\eeq
Using (\ref{Gest1}), \rf{n16} and \rf{n17} we obtain
\beq{n22}
\tilde{C}G^0(x;n)\leq (1-x)^{-n/2}\left(\left(1- {c''x^{1/3}\over 
|\log x|^{2/3}}\right)^n
+x^2\right).
\eeq  
 Inserting $n= [2(c'')^{-1}x^{-1/3}|\log x|^{5/3}]$ into
this inequality yields
\beq{n23}
\tilde{C}G^0(x;[ 2(c'')^{-1}x^{-1/3}|\log x|^{5/3}  ])\,\leq\, 3x^2\,
\eeq
for $x$ sufficiently small.
Finally, combining 
\rf{n23} with \rf{w16}, it follows that
\beq{n26}
e^{-nm(x)}\leq 3\tilde{C}^{-1}x^{3/2}
\eeq
 implying the lower bound in (\ref{mest}) for $x$ small enough.
 \hfill$\square$
\bigskip

\section{Discussion}
It is clear from the arguments in this paper that the existence of a unique
spine played a fundamental role.  This allowed us to decompose the trees
into an infinite line with identically distributed random outgrowths.  We
believe that our methods will allow one to calculate the spectral dimension
of any random tree ensemble with this property.  However, nongeneric trees
(sometimes called exotic trees) are not expected to have a unique
spine in general and it would be interesting to generalize the results of
this paper for such trees.  For results in this direction, see 
\cite{jwc,burda}.

Our results are all about ensemble averages.  Working slightly harder and
using the techniques of \cite{BarKum} we believe that one can obtain
estimates of the spectral dimension of generic trees valid with probability
one.

In the case of combs \cite{DJW} we saw that mean field theory holds in the
sense that the singularity of the ensemble average of the 
generating function for first return probabilities 
was given by that of $Q(x)$, i.e., 
$Q(x)\sim (1-P(x))^{-1}$.
Whether
this holds for trees remains to be seen. 

Clearly an outstanding problem is to discover the relevance of loops for the
spectral dimension of random graphs.  We only expect loops to
play a role in determining the spectral dimension 
if they in a suitable sense 
bound a large part of the graph.  It remains to turn this intuition into a
precise statement.  
Any graph can of course be made a
connected tree by removing links and it is not hard to see that cutting links
cannot increase the spectral dimension but may decrease it.    
The role of loops 
in determining the spectral dimension seems 
to be the crucial feature that one must understand in
order to get a grip on the spectral dimension of planar graphs and higher
dimensional random triangulations of interest in quantum gravity.

\bigskip

\noindent
{\bf Acknowledgement.}  This work is supported in part by MaPhySto funded
by the Danish National Research Foundation, Marie Curie grant
MRTN-CT-2004-005616, the Icelandic Science Fund and UK PPARC grant
PP/D00036X/1.

\bigskip

\noindent
\section*{Appendix A}
In this appendix we provide details of the proof of Theorem 2. It is
implicitly assumed below that $Z_N\neq 0$ for all
$N$, but the reader can easily verify that this is not an essential
limitation of the arguments. 

As explained in \cite{bergfinnur} it is sufficient to prove,
for arbitrary fixed value of $R\geq 1$, that 
\beq{convI}
\nu_N(\{\tau\in\Gamma : |B_R(\tau )| > K\})\,
\to\, 0 \quad\text{as $K\to\infty$}
\eeq
uniformly in $N$, and that the sequence
\beq{convII}
\left(\nu_N(\{\tau\in\Gamma : B_R(\tau)=\tau_0\})\right)_{N\in\mathbb N}
\eeq
is convergent for each finite tree $\tau_0\in\Gamma$.

We prove (\ref{convI}) by induction on $R$, the case $R=1$ being
trivial. Consider first the case  $R=2$.  If the volume of $B_2(\tau )=k+1$
then the order of the vertex $s_1$ next to the root is $k+1$.  
Letting $N_i$, $i=1,2,\ldots ,k$, denote the volumes of the 
various subtrees attached to $s_1$ we
have the following estimate,
\begin{eqnarray}
\nu_N(\{\tau\in\Gamma : |B_2 (\tau )|= 
k+1\})&=& Z_N^{-1}w_{k+1}\sum_{N_1+\dots+N_k=N-1}
\prod_{i=1}^k Z_{N_i}\nonumber\\
&\leq& k\,\zeta _0w_{k+1}\sum_{{N_1+\dots+N_k=N-1} \atop {N_1\geq
(N-1)/k}}\frac{Z_{N_1}\zeta _0^{N_1}}{Z_N \zeta _0^N}\prod_{i=2}^k
\left(Z_{N_i}\zeta_0^{N_i}\right)\nonumber\\
&\leq& C\,k^{5/2}w_{k+1}Z_0^{k-1}\label{R=2}
\end{eqnarray}
for $k\geq 1$, where we have used (\ref{u11}), and $C>0$ is a
constant independent of $k$ and $N$. Since $Z_0<\rho$ the last
expression tends to $0$ as $k\to\infty$, proving (\ref{convI}) for
$R=2$. Note that this inequality holds also for $k=0$ if $k^{5/2}$
is replaced by $(k+1)^{5/2}$.  

Assume (\ref{convI}) holds for some $R\geq 2$. Since the set of
different balls
$B_R(\tau)$ of volume at most $K$ is finite for each fixed $K>0$, it
suffices to show that 
\beq{convIII}
\nu_N(\{\tau\,:\,|B_{R+1}(\tau)| > K\,,\;
B_R(\tau)=\tau_0\})\,\to\, 0 \quad\text{as $K\to\infty$} 
\eeq
uniformly in $N$ for every finite tree $\tau_0$ of height $R$. Define
\beq{interim}
A_L(N,\tau_0) = \nu_N(\{\tau : |B_{R+1}(\tau)| = L\,,\;
B_R(\tau)=\tau_0\}).
\eeq
Let $M$ be the number of vertices in $\tau_0$ at distance             
$R$ from the root and let their orders be
$k_1, k_2,\ldots ,k_M$.
   Now we decompose any tree $\tau$ with $B_R(\tau
)=\tau_0$ into $\tau_0$ and subtrees whose root is at distance $R-1$ from
the root in $\tau$.  Note that the root link of these subtrees overlaps with
$\tau_0$, see Fig.\ 2.  
Then we have the following formula
\beq{uyt}
A_L(N,\tau _0) =Z_N^{-1}\sum_{{k_1+\dots k_M = L -|\tau_0|} \atop
{N_1+\dots + N_M
= N+M-|\tau_0|}}\left(\prod_{i=1}^M
Z_{N_i}^{(k_i)}\right)\prod_{i\in B_{R-1}(\tau_0)\setminus r}w_{\sigma_i}\,,
\eeq
where $Z_N^{(k)}$ is the contribution to $Z_N$ from trees
whose vertex next to the root has order $k+1$, that is
\begin{figure}[h!]
\begin{center}
\psfrag{tau}{$\tau_0$}
\includegraphics{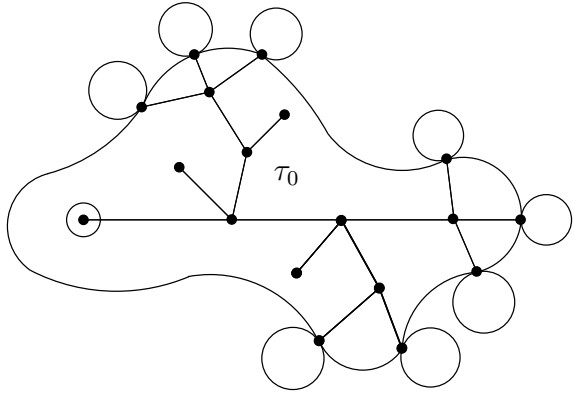}
\caption{The tree $\tau_0$ in the case $R=4$.  The circles denote finite
(possibly empty) trees that are attached to the vertices at distance $R$.}
\label{fig2}
\end{center}
\end{figure}
\beq{ytr}
Z_N^{(k)} = Z_N\,\nu_N(\{\tau :|B_2(\tau)|= k+1\})\,.
\eeq
Denoting the last product in (\ref{uyt}) by $W(\tau_0)$ and using
(\ref{R=2}) and \rf{ytr} we get
\beq{y66}
 A_L(N,\tau_0)\,\leq\,
W(\tau_0)\sum_{k_1+\dots+k_M \atop =L-|\tau_0|}\left(\prod_{i=1}^M
C(k_i+1)^{5/2}w_{k_i+1}Z_0^{k_i-1}\right) \sum_{N_1+\dots+N_M
\atop =N+M-|\tau_0|}\left(Z_N^{-1}\prod_{j=1}^M Z_{N_j}\right)\,. 
\eeq
The last sum can be estimated as in (\ref{R=2}) with the result
\beq{yy67}
 A_L(N,\tau_0)\,\leq\,
W(\tau _0)C^{M+1}M^{5/2}\,Z_0^{M-1}\sum_{k_1+\dots+k_M=L-|\tau_0|}\,\prod_{i=1}^M
(k_i+1)^{5/2}w_{k_i+1} Z_0^{k_i-1}\,,
\eeq 
provided $N$ is large enough.
Summing over
$L$ from $K+1$ to $\infty$ we obtain
\begin{eqnarray}\label{convIIII}
&\;&\nu_N(\{|B_{R+1}(\tau)| > K\,,\;
B_R(\tau)=\tau_0\})\nonumber\\&\leq&\!
W(\tau _0) C^{M+1}M^{5/2}\,Z_0^{-1}\sum_{k_1+\dots+k_M >  
K-|\tau _0|}\;\prod_{i=1}^M
(k_i+1)^{5/2}w_{k_i+1} Z_0^{k_i}\nonumber\\
\!\!\!&\leq&\! W(\tau _0) C^{M+1}M^{7/2}\,Z_0^{-1}\left(\sum_{k=1}^\infty
k^{5/2}w_{k} Z_0^{k-1}\right)^{M-1}\!\left(\sum_{k >
(K-|\tau _0|)/M}k^{5/2}w_kZ_0^{k-1}\!\right)\,.\nonumber\\
& &
\end{eqnarray}  
We have generic trees so the sum $\sum_{k=1}^\infty k^{5/2}w_{k} Z_0^{k-1}$
converges and (\ref{convIII}) follows from (\ref{convIIII}) 
since the last term in \rf{convIIII} tends to $0$ as $K\to \infty$.

It remains to verify (\ref{convII}). Summing over $L$ in (\ref{uyt})
we get 
\begin{eqnarray}\label{interim2}
\nu_N(\{B_R(\tau)=\tau_0\}) &=&
W(\tau _0)Z_N^{-1}\sum_{{k_1,\dots, k_M\geq 0}  \atop
{N_1+\dots + N_M
= N+M-|\tau_0|}}\prod_{i=1}^M
Z_{N_i}^{(k_i)}\nonumber\\
&=& W(\tau _0)Z_N^{-1}\sum_{
N_1+\dots + N_M = N+M-|\tau_0|}\prod_{i=1}^M
Z_{N_i}.
\end{eqnarray}
Now choose a constant $A$.  
By arguments identical to those of \cite{bergfinnur} p.\ 4804 one shows
that the contribution to the sum in \rf{interim2} from terms where $N_i\geq
(N+M-|\tau_0|)/M$ and $N_j \geq A$ for some pair of indices 
$i\neq j$ can be estimated
from above by $constant\cdot A^{-1/2}$ uniformly in $N$.  Note that the
condition on $N_i$ is always satisfied for at least one $i$.  The
remaining contribution is, for sufficiently large $N$,
\beq{convV}
W(\tau _0) \sum_{i=1}^M \sum_{N_1+\dots+N_M=N+M-|\tau_0| \atop N_j\leq A,\, j\neq
i}Z_N^{-1}\prod_{i=1}^M Z_{N_i}.
\eeq
Letting $N\to\infty$ with fixed $A$, the last expression converges to
\beq{yy77}
MW(\tau _0)\left(\sum_{N=1}^A
Z_N\zeta _0^N\right)^{M-1}\zeta _0^{|\tau_0|-M},
\eeq
by Lemma 1.
Letting $A\to\infty$ we conclude that 
\beq{convVI}
\nu_N(\{B_R(\tau)=\tau_0\})\underset{N\to\infty}\longrightarrow M W(\tau _0)
Z_0^{M-1} \zeta _0^{|\tau_0|-M}\,,
\eeq 
which proves (\ref{convII}).

Note that the above estimates show that for any constant $A$, all the
$N_j$'s except one are bounded by $A$, with a probability which tends to $1$
as $A\to\infty$, while one of them gets very large when
$N$ gets very large. 
A slight generalisation of the arguments leading to \rf{convVI} (see
\cite{bergfinnur}) shows that $\nu$ is concentrated on the set of
infinite trees with a single spine and that $\nu$ can be characterised
as explained in Section 2.

\bigskip 

\noindent
\section*{Appendix B}
\bigskip

In this appendix we establish the inequality (\ref{Cset}), that is
\beq{b11}
\nu(\{\tau : 
|B_R(\tau )|< \lambda\,R^2\}) \leq e^{-c_0\lambda^{-1/2}}\,,\quad\text{for
$R>0$ and \,$0<\lambda<\lambda_0$}\,, 
\eeq
where $c_0$ and $\lambda_0$ are positive constants.

Given $\tau\in\Gamma$ with one infinite spine, let $B^i_R(\tau)$
denote the intersection of the ball of radius $R$, centered at the
$i$th spine vertex $s_i$, with the finite branches attached to the spine at
$s_i$. 
Since 
\beq{ppp}
B_{[R/2]}^1(\tau)\cup \dots \cup B_{[R/2]}^{[R/2]}(\tau)\,\subseteq\, B_R(\tau) \,,
\eeq
it is sufficient to prove
\beq{Ysum}
\nu(\{\tau : |B_R^{1}|+\dots +|B_R^R| < \lambda\,R^2\}) \leq
e^{-c_0\,\lambda^{-1/2}}\,. 
\eeq
Let us fix $i\in\mathbb N$ and set $|B_R^i| = Y_R$.
Then 
\beq{b12}
Y_R =  X_1+\dots + X_R\,,
\eeq
where $X_n(\tau)$ is the number of vertices in $B_n^i(\tau)$ at
distance $n$ from $s_i$.

From the properties of $\nu$ given in Section 2 and the discussion of 
Section 3 
it follows that the
generating functions 
\beq{b13}
h_n(z) = \sum_{a=0}^\infty \nu ( \{ \tau : X_n(\tau )=a\})\,z^a 
\quad\text{and}\quad k_n(z)=\sum_{b=0}^\infty \mu (\{ T :D_n(T) =b\})\,z^b,
\eeq
with $D_n$ as in Section 3, are related by
\beq{b16}
h_n(z) = f'(k_n(z))\,.
\eeq  
Since 
$\nu(\{X_n > 0\}) = 1- h_n(0)$
and
$\mu(\{D_n > 0\}) = 1- k_n(0)$
it follows from (\ref{heightdistr}) that
\beq{Xtail}
\nu(\{X_n > 0\}) = \frac 2n + O(n^{-2})\,.
\eeq
Similarly, the generating functions 
\beq{b17}
g_R(z) = \sum_{a=0}^\infty \nu (\{\tau : Y_R=a\})\,z^a
\quad\text{and}\quad
f_R(z)=\sum_{b=1}^\infty \mu (\{T: |B_R(T)|=b\})\,z^b  
\eeq
 are related by
\beq{b18}
g_R(z) = f'(f_R(z))\,.
\eeq
Here $f_R$ is determined by the recursion relation
\beq{b19}
f_{R+1}(z) = zf(f_R(z))\,,\quad f_1(z) = z\,.
\eeq
Differentiating and evaluating at $z=1$, one easily deduces from these
two relations that
\begin{eqnarray}
\langle\, |B_R|\,\rangle_\mu &=& R\,,\label{muvol}\\
\langle\, Y_R\,\rangle_\nu &=& f''(1)R\,,\label{Yasymp}\\
\langle\, |B_R|^2\,\rangle_\mu &=& \frac 13 f''(1)R^3 + O(R^2)\,,\\
\langle\, Y_R^2\,\rangle_\nu &=& \frac 13 f''(1)^2 R^3 + O(R^2)\,.
\label{Y2asymp}
\end{eqnarray}
Using (\ref{Xtail}), (\ref{Yasymp}) and (\ref{Y2asymp}) we now prove,
following \cite{BarKum}, that 
\beq{Ygeq}
\nu(\{Y_R\geq c'R^2\})\; \geq\; \frac{c''}{R}\,,
\eeq
for suitable positive constants $c',c''$. For this purpose let ${\cal
A}_n =\{\tau : X_{[n/2]}>0\}$.  Then, by \rf{Xtail},
\beq{b20}
\nu({\cal A}_n) = \frac 4n + O(n^{-2})\,.
\eeq
Since $Y_n = Y_{[n/2]}$ on the complement of ${\cal A}_n$, we have by
\rf{Yasymp},
\beq{b21}
f''(1)(n-[n/2]) = \int_{{\cal A}_n}(Y_n-Y_{[n/2]})\,d\nu \leq \int_{{\cal
A}_n}Y_n \,d\nu\,,
\eeq
and therefore
\beq{b23}
\langle\, Y_n\,|{\cal A}_n\rangle_\nu\, \geq\, \frac{f''(1)n}{2\nu({\cal A}_n)}\,
=\, \frac{f''(1)n^2}{8} + O(n)\,\geq\, c_1 n^2\,,
\eeq
where $\langle\,\cdot\,|{\cal A}_n\rangle_\nu$ denotes the expectation
w.r.t.\ $\nu$ conditioned on ${\cal A}_n$. We also have 
\beq{b24}
\langle\, Y_n^2\,|{\cal A}_n\rangle_\nu\; \leq\; \frac{\langle
Y_n^2\rangle_\nu}{\nu({\cal A}_n)}\, =  
\frac{1}{12}\,f''(1)^2 n^4 + O(n^3)\,\leq\,c_2 n^4\,,
\eeq
by \rf{Y2asymp} and \rf{b20}.
Combining the last two inequalities with the reversed Chebyshev
inequality for the conditional expectation $\langle\,\cdot\,|{\cal
A}_n\rangle_\nu$,
\beq{revCheb}
\nu({\cal A}_n)^{-1}\nu(\{Y_n \geq \frac 12 \langle Y_n|{\cal A}_n\rangle_\nu\}\cap{\cal
A}_n)\geq \frac{\langle Y_n|{\cal A}_n\rangle_\nu^2}{4\langle
Y_n^2|{\cal A}_n\rangle_\nu}\,, 
\eeq
see e.g.\ \cite{feller} p.\ 152, we obtain 
\beq{b25}
\nu(\{Y_n\geq\frac 12 c_1 n^2\})\;\geq\; \nu(\{Y_n \geq \frac 12 c_1
n^2\}\cap {\cal A}_n)\;\geq\; 
\frac{c_1^2}{4c_2}\nu({\cal A}_n)\, \geq\,  \frac{c''}{n}\,,
\eeq
which proves (\ref{Ygeq}).

Finally, (\ref{Ysum}) can be obtained from (\ref{Ygeq}) as
follows. Given $\lambda\in (0,c')$ let $n=[(\lambda/c')^{1/2}R]$. Then
$n<R$ and
\begin{eqnarray}
\{|B_R^1|+\dots + |B_R^R|\leq \lambda R^2\}&\subseteq& 
\{|B_n^1|+\dots + |B_n^R|\leq \lambda R^2\}\nonumber\\
&\subseteq& \{|B_n^1|+\dots + |B_n^R|\leq  c'n^2\}\nonumber\\
&\subseteq& \{|B_n^i|\leq c' n^2\,,\;i=1,\dots, R\}\,.
\end{eqnarray}
Hence, using (\ref{Ygeq}) and noting that 
$|B_n^i|\,,\,i=1,\dots,R$, are independent                            
random variables, we obtain
\beq{b33}
\nu(\{|B_R^1|+\dots + |B_R^R|\leq \lambda R^2\})\,\leq\,
\left(1-\frac{c''}{n}\right)^R\, \leq\,
\left(1-\frac{c'^{1/2}c''\lambda^{-1/2}}{R}\right)^R\,. 
\eeq
This proves (\ref{Ysum}) with $c_0=c'^{1/2}c''$ and 
$\lambda _0=c'$.

\end{document}